\documentclass[preprint2]{aastex}

\shorttitle{Cosmic silicate glass optical functions}
\shortauthors{Speck et al.}

\begin{document}

\title{Better Alternative to ``Astronomical Silicate'': 
     Laboratory-Based Optical Functions of Chondritic/Solar \\ 
     Abundance Glass With Application to HD 161796}

\author{A. K. Speck}
\affil{Department of Physics and Astronomy, University of Missouri-Columbia, 
Columbia, MO 65211, USA}
\email{speckan@missouri.edu}


\author{K. M. Pitman}
\affil{Planetary Science Institute, Tucson, AZ 85719, USA}
\affil{{\rm now at} Space Science Institute, Boulder, CO 80301, USA.}

\and

\author{A. M. Hofmeister}
\affil{Department of Earth and Planetary Sciences, Washington University, 
St. Louis, MO 63130, USA}

\begin{abstract}

``Astronomical'' or ``circumstellar'' silicate optical functions 
(real and imaginary indices of refraction $n(\lambda)$ and $k(\lambda)$)
have been previously derived from compositionally and structurally 
disparate samples; past values were compiled from different sources in the 
literature, and are essentially kluges of observational, laboratory, 
and extrapolated or interpolated values. These synthetic optical functions 
were created because astronomers lack the 
quantitative data on amorphous silicates at all wavelengths needed
for radiative transfer modeling.  
This paper provides optical functions that (1) are created with a consistent 
methodology, (2) use the same sample across
all wavelengths, and (3) minimize interpolation and extrapolation 
wherever possible. 
We present electronic data tables of optical functions derived 
from mid-ultraviolet to 
far-infrared laboratory transmission spectra 
for two materials: iron-free glass with chondritic/solar atmospheric 
abundances, and metallic iron.
We compare these optical functions to 
other popular $n$, $k$ data used to model amorphous silicates 
(e.g., ``astronomical'' or ``circumstellar'' silicate), 
both directly and in application to a simple system: the dust shell of the 
post-AGB star HD\,161796.
Using the new optical functions, we find that the far-IR profile of 
model SEDs are significantly affected by the ratio of glass to iron.
Our case study on HD\.161796 shows that modeling with our new optical 
functions, the mineralogy is markedly different from that derived using 
synthetic optical functions and suggests a new scenario of crystalline 
silicate formation.

\end{abstract}


\section{Introduction}
\label{intro}

Silicate grains dominate the emission and extinction processes in many astrophysical 
environments but astronomers still do not know exactly what composition or structure 
constitutes these 
dust grains or how these properties change from location to location. 
Astromineralogy, the study of the precise and detailed nature of dust grains in space, 
has developed rapidly over the past few decades 
\citep[see reviews in][]{speck97,speck98,speck00,Molster00,WatersMolster99,Henning03,Kwok04,Pitman_etal10,GuhaNiyogi, Speck12}.
However, interpretation of dust grains is still limited by the paucity and quality of mineral 
data used for comparison with and modeling of astronomical observations.

While dust in space is particulate in nature, there is some discussion about whether it is 
appropriate to use bulk materials for spectroscopy of candidate solids \citep[e.g.][and references therein]{Speck99,Henning03}.
It has already been shown that thin films (slabs) and dilute particle distributions 
result in the 
same spectral features \citep{Speck99}. Moreover, having optical constants from 
spectra of bulk materials is important because they set an upper limit for extinction through 
a certain material \citep{hofm09}.  

Although laboratory measurements of optical functions for astronomical applications 
are of high value, they characterize individual analogue materials and are not the real dust 
we see in space. However, good quality extinction properties of the components 
of the real space dust can be derived from these analogue materials. Likewise, deriving a set of 
effective optical functions from astronomical data would be acceptable if the 
temperature and density structure, and grain size and shape distribution of a system was  
well known. Here we aim to provide optical functions that allow us to determine these 
details in astronomical environments more completely.

There are two basic approaches to identifying dust minerals in space:
  (1) Work backward from astronomical observations where temperature and density structure of a 
system is well known to derive effective optical functions that approximate those of real dust 
in space.
  (2) Work forward from laboratory measurements of dust analogue minerals whose compositions and 
impurities are well characterized to derive optical functions of minerals predicted to occur in 
space.

Ideally, we should use both approaches and find where (1) and (2) match, however, astronomers 
do not take the dual approach.
 Approach (1) alone does not reveal anything about the chemistry; those optical functions are 
not useful for diagnosing changes from location to location and should not be used to 
interpret compositional effects elsewhere.  
We are taking approach (2) by providing dust analogue data that will 
provide insight into the chemistry. Moreover we are correcting past efforts which made
kluges that hybridize rather than compare the two approaches.

The simplest approach for determining the nature of dust grains in space is 
to match the positions, widths, and strengths of observed spectral features with those 
seen in laboratory spectra. This is usually achieved by fitting a continuum 
to the observed spectrum and dividing (or subtracting) 
to derive emissivities of the observed spectral features. 
These emissivities are then compared to absorption spectra\footnote{%
Typical absorption spectra are derived from transmission experiments which yield the 
absorbance $a$ or the absorptivity $A$. The details of these different parameters and 
how they are derived are reviewed in \citet{Speck2014}.}
 produced in the laboratory.
Sometimes the emission/absorption profile is calculated from optical functions 
\citep[complex refractive indices; $n_\lambda$ and $k_\lambda$; see e.g.][]{GuhaNiyogi}\footnote{%
The complex refractive indices of a substance are often referred to 
as the ``optical constants.'' Because these values are not constant, but rather 
 vary with wavelength, temperature, and polytype (structure) as well as chemical 
composition of the sample, we use the term ``optical functions'' in this work.}.
 Radiative transfer (RT) modeling uses the optical functions\footnote{%
while RT models use absorption and scattering, or extinction efficiencies of candidate 
dust grains, many are equipped to use optical functions as inputs and do the conversion 
themselves, then generate the required cross-sections/efficiencies. 
Providing optical functions allows for more versatility in the modelling 
(e.g.\ grain size and shape effects)}
minerals to predict how a given object should look both spectroscopically and in images 
\citep[e.g.][]{pinte08,eisner05,meixner,mensch02,voors00,sorrell}.
RT modeling can be used to determine the effects of grain size distributions 
and mineralogy on the expected spectrum, and places constraints on the 
relative abundances of different grain types in a dust shell. 
However, optical functions used in such RT models must cover a broad wavelength range, 
from the ultraviolet (UV) to the infrared (IR).
Deriving optical functions over this broad range requires laboratory transmission or 
reflectance spectra from multiple detectors; this is one of many reasons that $n,k$ 
are hard to find.
%
As a result, various compilations have been produced that kluge together observed and 
laboratory-collected data in order to produce optical functions to fill the full 
wavelength grid
required by RT codes.

Still the most widely used optical functions for silicates are those from 
\citet*{dl84} \citep[hereafter]{dl84}, \citet{draine85, draine03} and \citet*{ohm92}
\citep[hereafter]{ohm92}
which were originally generated 30 years ago and included 
kluges out of necessity. However, we are now able to provide optical functions 
which are kluge-free from the far-IR to the UV ($0.19 < \lambda < 250$\,$\mu$m).
Here we present new laboratory-generated complex refractive indices
for an iron-free glassy silicate of chondritic/solar atmospheric composition 
using a single sample for all wavelength regimes and compare it to the most  
commonly used ``astronomical silicate'' optical functions. 
We apply our new silicate glass optical functions to modeling the 
dust around HD\,161796, an oxygen-rich pre-planetary nebula, in order to compare our new
optical functions to those most commonly used.
In addition we present new optical functions for metallic iron in order to 
properly model the relevant astrophysical environments.

There~~ are~~ a~~ number~~ of~~ published~~ laboratory~~ datasets~~ on~~ metallic~~ iron,~~ 
(see also \\
http://www.astro.spbu.ru/JPDOC/f-dbase.html), however, 
most have a very limited wavelength range.
In astronomy, the most  widely used data are from \citet{ordal}  
which have both low resolution and 
do not extend to wavelengths shortward of the red-end of the visible spectrum. 
\citet{pollack94} also present a figure which contains a broad wavelength 
range optical function dataset for metallic iron, but this is compiled from multiple 
samples and thus not discussed further here. Likewise, 
there is a set of optical functions for metallic iron from \citet{lynch}
in \citet{Palik91}, 
but it also is derived from a compilation of samples.
Iron optical properties are also important for remote sensing in planetary science, but 
those studies are usually limited to the NUV--NIR range
\citep[see][whose own excellent data only covers 0.16--1.7\,$\mu$m]{cahill1}.

The new Fe optical functions improve upon the previous datasets 
because (a) most previous datasets do not extend beyond the red-end of the visible and 
(b) many are compilations from disparate samples. This leads to artifacts, such as a 
discontinuity in $R$ at 600\,nm. In contrast, our new data has
the higher spectral resolution, broader wavelength coverage, and is derived from the 
same sample through that wavelength range.

\section{Past work: ``Astronomical'' and ``circumstellar'' silicates}
\label{pastwork}

Several different synthetic optical functions were derived at a time when 
there was insufficient laboratory data to be able to model astrophysical environments 
across a broad wavelength regime. A detailed discussion of lab data for Si-based materials can be 
found in \citet{Colangeli2003} as well as \citet{SWH11}, 
and a wider discussion of cosmic silicates can be found in \citet{Henning2010}.  
\citet{SWH11} made detailed comparisons between previous published laboratory data on 
silicate glasses.

Compared to the wealth of data on potential astrominerals, our literature search 
\citep[e.g.\ http://www.astro.spbu.ru/JPDOC/f-dbase.html][and references therein]{Henning1999}  
revealed that although silicates have been studied in more detail than other compounds, 
there are still shortcomings in the available data with respect to silicate glasses.  
Specifically, existing lab data on glasses do not allow astronomical spectra to be 
interpreted reliably in terms of either M$^{2+}$/Si or Mg/Fe ratio, which are
important tools for discriminating between competing dust formation models 
\citep[c.f.][Figs. 7--9]{SWH11}.
\citet{Pitman2013} compared the UV-Visible extinction coefficients 
(and imaginary index of refraction; $k$) for a range of silicates, both amorphous and crystalline 
with those from a variety of previously published samples.
The figure from \citet{Pitman2013} (reproduced in Fig.~\ref{fig:pitman2013}) illustrates 
that $k$ values vary wildly for both crystalline and amorphous silicates in the UV-vis, whether 
the datasets are from the lab or from modelers. The 70\% and 80\% Mg-rich silicate glasses 
of \citet{dorschner95}, laser-ablated amorphous forsterite and enstatite from \citet{sd96}
and the ``astronomical silicates'' of \citet{dl84}, \citet{ohm92}, and \citet{dp95} 
have a similar upper limit 
on $k$. The $k$ values for the remaining \citet{dorschner95} Mg-rich silicate glasses, 
sample 1-S from \citet{jager94}, and enstatite and olivine from \citet{EH1975} are a 
factor of ten less. 
The datasets from {\citet{huffman75} are in a class by themselves 
(setting the uppermost limit on $k$), as are the Fe-rich silicate glasses. 
It is clear from that ostensibly 
similar compositions give rise to very different optical functions. 
In addition for crystallinity and composition differences, the variation in derived optical
functions is also due to the use of powders, rather than bulk samples. Studis involving powder
samples often overestimate the $k$ values because of internal reflections and scattering. 
This problem also afflicts bulk samples which include imperfections
\citep[see][for more detailed discussion of this issue.]{hofm09, Speck2014, Speck99}.

\begin{figure}[!h]
\includegraphics[angle=270, scale=.425]{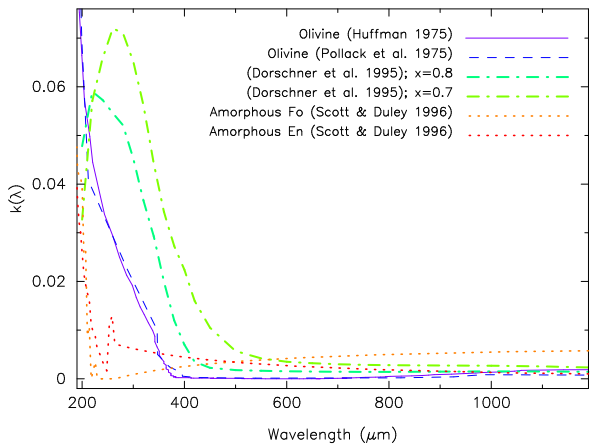}
\caption{NUV--NIR Imaginary index of refraction, $k$, (a.k.a. extinction) of commonly used 
``Astronomical'' silicates \citep[reproduced from][]{Pitman2013} \label{fig:pitman2013}}
\end{figure}

 
Here we focus our discussion on the most commonly used datasets for analysis 
of silicate dust \citep{dl84}, \citet{draine85, draine03} and \citet{ohm92}.
Other artificial ``circumstellar,'' or 
``astronomical'' silicates, based in part on real mineral data, 
also exist \citep{suh91,laor93,dp95}. 
The optical functions in 
\citet{suh91} 
were adopted from \citet{dl84} optical 
functions shortward of $\lambda=$8\,$\mu$m and are thus not discussed here.
\citet{dp95} followed a similar methodology to \citet{dl84}, 
but using the then new IRAS observations of $\sim$300 red supergiants. 
However, it is not widely used, probably because \citet{ohm92} had already 
supplanted \citet{dl84} for some purposes, and \citet{ohm92} incorporated the 
source data for \citet{dp95} \citep[from][]{dp90} into their discussion.
\citet{laor93} is mostly based on \citet{dl84}, but with tweaks at the 
high energy end ($>20$eV) which are calculated for an assumed density and
MgFeSiO$_4$ composition, but these modifications were supplanted by 
\citet{draine03} and not considered further here.
\citet{kimartin1} found a problem using \citet{dl84} to model UV polarization 
which they attributed to the use of crystalline olivine data in the 
derivation of ``astronomical silicate'' optical functions in \citet{dl84}.
Thus they proposed using measurements from volcanic glass in the UV to alleviate the problem. This is essentially still a kluge.
In light of the above, in this work, we will focus on \citet{dl84,draine03} and 
\citet{ohm92} datasets in our discussion.
\\
%

Artificial ``cosmic,'' ``circumstellar,'' 
``interstellar,'' or ``astronomical'' silicate, optical and dielectric 
functions are not optimized at all wavelengths.  For example, the 
``astronomical silicate'' from \citet{dl84} is optimized for 
the mid-IR, but not in the UV-vis range \citep{pollack94}.
\citet{sorrell} found that while \citet{dl84} ``astronomical silicate'' could 
produce a satisfactory match to the 10\,$\mu$m feature in a protostellar object, it 
could not simultaneously match the UV region \citep[c.f.][]{kimartin2}.

Synthetic ``astronomical'' or ``circumstellar silicate'' optical functions are 
often favored because they have broad wavelength coverage, especially at short 
wavelengths (UV and X-ray).  As stated earlier, 
optical functions derived from merged laboratory 
spectra are collected piecemeal using multiple detectors; 
merging must be performed judiciously by the 
user. Furthermore, laboratory derived $n$ and $k$ values typically do not 
cover very far into the UV and X-ray 
range because these detectors must be hand built 
or, if commercially available, are very expensive. 
However, the trade-off in using synthetic ``astronomical'' 
or ``circumstellar'' silicate optical functions is that chemical composition has 
not been preserved across all wavelengths.  
The optical functions currently in widespread use were constructed by a 
combination of forward calculation from laboratory spectra of chemically disparate 
samples and backward calculation from
astronomically-observed dust opacities, together with pieces that are 
theoretically derived as illustrated in  Figure~\ref{ohm_dl}.
Those datasets also include modifications to specifically match certain 
astronomical observations.  Consequently, whereas the synthetic optical 
functions match some interstellar spectra 
\citep[in the case of][]{dl84, draine03} 
or circumstellar spectra \citep[for][]{ohm92}, those $n$ and $k$ values are not useful 
for diagnosing changes from location to location and should not be used to 
interpret compositional effects.  Indeed, \citet{dl84}, \citet{draine03}, 
and \citet{ohm92} explicitly stated such caveats, but their warnings have been largely 
ignored because better alternatives have not been available.

\begin{figure}[!h]
\includegraphics[angle=270, scale=.35]{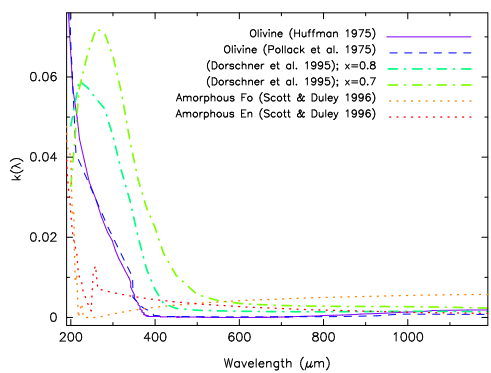}
\caption{Source data for different wavelength regimes in previous 
``astronomical silicate'' optical functions from \citet{dl84}.
Solid lines are the $n$ values.
dashed lines are the $k$ values.
These $n$ and $k$ are merged from a mix of experimental, observational, and theoretical data or assumptions.
Main figure has linear axes: $x$-axis is wavelength in $\mu$m; 
$y$-axis is the value of the real or imaginary refractive index.
Inset is the same as main figure but with log-log axes. \label{ohm_dl}}
\end{figure}


\citet{dl84}\footnote{Electronic tables of the dielectric function 
($\epsilon$) and equations to convert these to $n,k$ were published in 
\citet{draine85}.} produced synthetic optical functions for both ``astronomical'' 
silicate and graphite in order to model interstellar dust. Because of the 
paucity of laboratory data at that time,
\citet{dl84} compiled data from multiple 
sources in order to achieve continuous wavelength coverage.
The full procedure is described by \citet{kimartin1} and summarized in Fig.~\ref{ohm_dl}

The updates by
\citet{draine03} apply mostly to the short wavelength/high energy end of the 
spectrum such that the values from \citet{dl84} were slightly modified:
(1) a feature was removed at $\lambda^{-1}$ = 6.5\,$\mu$m$^{-1}$ (0.15\,$\mu$m); 
(2) for the sub-mm range (250\,--\,1100$\mu$m) the value of $\epsilon_2$ was varied by
$\pm$12\% following \citet{ld01};
(3) above 30\,eV ($\lambda <$ 40\,nm) $\epsilon_2$ was estimated from atomic 
photo-absorption cross-sections, 
except near absorption edges. 
These high energy  dielectric function values were merged with original functions from 
\citet{draine85} between 18 and 30\,eV 
(40$< \lambda <$70\,nm).

\citet{ohm92} followed a similar method to that of \citet{dl84} and \citet{draine03} but 
with different assumptions, constraints and observational data.
In particular, \citet{ohm92} used an average of mid-IR observations from IRAS 
compiled by \citet{vk88}. As in \citet{dl84} and \citet{draine03}, it was assumed that 
grains have an MRN grain-size distribution\footnote{%
i.e., the number of grains of size $a$ is given by $n(a)$ proportional 
to $a^{-q}$, where $n$ is the number of the grains in the size interval 
$(a,a + da)$ and $q=3.5$;$a_{min}=0.005\mu$m; and $a_{max}=0.25\mu$m 
\citep{Mathis1977}.}.  
They also assumed a continuous distribution 
of ellipsoid shapes. For the NIR-MIR merge region, \citet{ohm92} used 
laboratory data from magnetite (Fe$_3$O$_4$) and metallic iron 
to get high enough opacities.

Both ``astronomical'' and ``circumstellar'' silicates 
will match some spectral features, and can be used for comparing optical depths 
between different dusty environments.
However, because those $n$ and $k$ values are derived from a combination of 
laboratory and observed data as described above, they do not 
represent real solids across all wavelengths
and thus cannot be used to determine the physical and 
chemical properties of dust in space, or how and why dust varies spatially or temporally.
Moreover, because in the majority of systems, 
most light is absorbed and scattered in the 
NUV-NIR region, calculations of dust abundances are based primarily on 
optical properties for species that are expected to be rare
( e.g., crystalline olivine\footnote{%
based on the interpretation of observations})
or are approximations (e.g., alumina/corundum Al$_2$O$_3$). 
In addition, because metallic iron is invoked either 
directly \citep[in the case of][]{ohm92} or 
indirectly \citep[using ``dirty silicates'' in the 
case of][]{draine03},  the effect of iron abundances on 
dust opacities cannot be determined with certainty. 
Because iron is predominantly generated in Type~Ia supernovae and 
the $\alpha$-rich silicon and magnesium atoms are predominantly made in 
core-collapse (Type II) supernova \citep{bulbul,iwamoto,nomoto}, 
the relative abundances of these 
dust-forming elements may not scale together as we move from place to place, or 
especially back in time. The separation of iron is essential to understanding 
chemical evolution.

Finally, while optical functions do not depend on grain size/shape; 
the resulting absorption and scattering properties do. 
Consequently, the derivation of optical functions from 
(absorption and scattering) observations necessarily 
requires assumptions about the grains, in particular their size and shape 
distributions.
 For all the synthetic optical functions above, the grains are 
assumed to have MRN size distributions. 
Furthermore, \citet{dl84} and \citet{draine03} assumed a specific form of oblate grains, 
while \citet{ohm92} assumed a broad continuous distribution of ellipsoids
(including spheres, pancakes and needles). 
%

\section{Alternative: ``cosmic silicate'' glass}

Here we present new optical functions for a ``cosmic silicate'' glass with 
chondritic/solar atmospheric abundances of the major elements excluding iron. 
These new optical functions were derived from laboratory spectra 
using the same sample of silicate glass 
from $\lambda$=0.19 to 250\,$\mu$m. 
The composition is given in Table~\ref{tab:composition}.
As described in \citet{SWH11}, we excluded iron from the sample for two reasons: 
(1) to better preserve oxidation state; 
and (2) because several theoretical models of dust condensation suggest that 
iron does not commonly form silicates, but rather that it forms metallic iron 
grains \citep[e.g.,][]{lf99}. 
Iron atoms are almost as common as magnesium and silicon atoms 
in the solar neighborhood, leading to some debate about the inclusion of iron 
in silicates. However, compositions that include grains that are 
agglomerations of iron-free silicates and metallic iron pieces are commonly 
invoked. Indeed, ``dirty silicates'' were suggested because iron-free silicates 
are too transparent \citep{Merrill76}.

In addition to our $n$ and $k$ for ``cosmic silicate'' glass, we also present
$n$ and $k$ for metallic iron in order to demonstrate this opacity issue. 
Because our silicate is Fe-free, and we provide Fe optical functions, 
astrophysicists can now confidently investigate the effect of changing Mg/Fe 
abundance ratios in environments where dust forms.
This is especially important in lower 
metallicity environments, where Mg/Fe should be higher than solar even 
though the absolute number density of atoms may be lower.

\begin{center}
\begin{table}[!h]
\caption{Composition of ``Cosmic Silicate'' glass:  
\underline{$\rm (Na_{0.11}Ca_{0.12}Mg_{1.86})(Al_{0.18}Si_{1.85})O_6 $ }}
\label{tab:composition}
\begin{tabular}{lr}
\hline											
oxide	&	weight percent  \\
\hline
  SiO$_2$	&	54.26	\\
 Al$_2$O$_3$ 	&	4.34	\\
 MgO 	&	36.58	\\
 CaO 	&	3.27	\\
 Na$_2$O 	&	1.61	\\
total 	&	100.05	\\
\hline
water   &  ppm  \\
content &       \\
H$_2$O	&  81	\\
\hline
\end{tabular}
\end{table}
\end{center}

\section{Experimental methods}
\label{sec:expmeth}

\subsection{Sample preparation}

\subsubsection{Glass synthesis}
\label{makegl}

The sample studied here, ``cosmic silicate,'' is an Fe-free silicate glass with 
chondritic/solar relative
Mg, Al, Si, Ca, and Na abundances, synthesized as described by 
\citet{getson07} and \citet{SWH11}
by A. Whittington at the University of Missouri in Columbia.  
%
%
%
%
The melting and quenching techniques used produce more consistent, 
fully amorphized glasses than other sample preparation techniques 
\citep[e.g., 
chemical vapor deposition, ion-irradiation, laser ablation; see discussion by][]{SWH11}. 
The ``cosmic silicate'' glass is free of bubbles and internal crystals; electron 
microprobe composition from \citet{SWH11}
is given in Table~\ref{tab:composition}. 
%

Our ``cosmic silicate'' glass has a trace amount (81 ppm) of H$_2$O, 
and shows up as a very weak feature around 3\,$\mu$m.  
For a binary, ternary, or quaternary silicate glass, we would expect to see at 
least one absorption in the range of 2.7 to 4.0\,$\mu$m, 
depending on the composition of the glass \citep{brown}.  
Given that $\epsilon '$ is proportional to constant $+ 10^{-22}$ times the concentration at 
long wavelengths \citep[see e.g.][]{Andeen}, we can safely neglect the trace H$_2$O 
amount in the IR.

\subsubsection{Metallic iron}
High purity electrolytic iron ($>$99.97\%) was purchased from 
Alfa/Aesar\textregistered in the form of $\sim$cm-sized slabs with 
$\sim$mm thicknesses.

\subsubsection{Sample thickness determination}

We acquired spectral measurements of both the ``cosmic silicate'' 
glass and the metallic iron sample from polished slabs (in the UV-VIS-IR) 
and thin films (in the IR only).  
All polished slab samples were cut to sub-mm scale with a 
rock saw, then progressively ground, polished with micron-sized grits by hand, and 
(for transmission measurements in the IR) were thinned by compression in a 
diamond anvil cell to films of below 1\,$\mu$m thick to 
achieve well-resolved spectral peaks and good signal-to-noise ratios 
in the spectra.  

Many published papers and \citet{bh83}
suggest that grain shape has an effect on absorption and scattering 
cross-sections.
Measuring optical properties from slabs, rather than loose or 
dispersed powders, ensures that the optical properties
depend only on composition and not 
on grain shape or size.  Also, using samples with differing thicknesses 
for different spectral segments helps to minimize the effect of back reflections
\citep{hofm09,Speck2014}.

For the thin films, the thickness of the film is the 
major driver of spectral quality.  For the polished slabs, the quality of 
the laboratory spectra is affected by the quality of the sample's polish 
(for reflectivity) and the thickness of the sample (millimeter pathlength 
for reflectivity, lower limit of $\sim$40\,$\mu$m achievable in the 
lab for transmission/absorption).  We polished the slab samples to a 
surface roughness that is equivalent to that of the spectrometer's reference 
mirror for the IR instrument. In the UV, 
the high absorbance overwhelms the small effect of scattering from the surface.

The thickness of the sample, $L$, which factors into the determination of 
$k$, was measured for the polished slabs using a digital micrometer or a 
calibrated reticule in a doubly polished microscope.  If spectral fringes 
were observed, we calculated the slab thickness from Eq.~\ref{slabthick},

\begin{equation}
L = ( 2 n_{vis} \Delta \nu )^{-1},
\label{slabthick}
\end{equation}

\noindent
where $\nu$ is frequency in cm$^{-1}$ and $n_{vis}$ is a tiepoint for 
the real index of refraction in the visible.  Our best determination of 
$n_{vis}$ of a thin ($\sim$mm-scale), doubly polished slab of ``cosmic 
silicate'' glass using a gemological refractometer is 1.60$\pm$0.01, which 
is close to values for other glasses with like density. 
For the ultrathin 
slabs, interference fringes occur near the Christiansen minimum;\footnote{%
The Christiansen minimum occurs where the refractive index of a medium 
changes rapidly and may approach that of the surrounding medium 
(usually air or vacuum). This results in very little reflection or absorption 
\citep{Conel69}.}
below, we use 
reflectivity data to compute $n$ and ultrathin slab thickness.  For the 
thin films, because the fringes are in the transparent region when $n =$1.6, 
we instead recorded the spacing of the diamond faces to estimate the film 
thickness.  Spectra were aquired for an ultrathin slab with $L =$38\,$\mu$m, 
two thick slabs with $L =$0.47 mm and 2.364 mm, and two micron-scale thin 
films.  Using $n =$1.6 gave 7.8\,$\mu$m for the thick film and 3.5\,$\mu$m 
for the thin film.  Later spectral scaling suggests that $L =$1.78\,$\mu$m 
for the thinner film.  The thicker of the two thin films was double the 
absorbance of the thin film, which is also consistent with the fringes.

\subsection{Laboratory spectroscopy}

We used several laboratory detectors at Washington University in St. Louis 
to measure room temperature (18--19$^{\circ}$C) spectra across the UV-IR 
wavelength range.  In the UV-VIS ($\lambda =$190--1100 nm, 
$\lambda =$0.19--1.1\,$\mu$m, or $\nu =$52,630--9090 cm$^{-1}$), we 
acquired unpolarized transmission and specular reflectivity spectra using 
the double-beam UV-1800 Shimadzu\textregistered UV-VIS Spectrophotometer 
to probe the mid-UV and the UV tail at high resolution (1 nm).  The UV-VIS 
spectra were not acquired in an evacuated chamber.

The infrared spectrum for ``cosmic silicate'' glass was previously published in
\citet{SWH11}. The same methodology from \citet{SWH11} and \citet{hofmeist03}
was applied to measure the IR spectrum for metallic iron in this work.

\subsubsection{Spectral merging}

For the ``cosmic silicate'' glass, spectral segments acquired for the 
two thin films, the ultrathin slab, and the two thick slabs were scaled 
in the region of wavelength overlap and merged.  When scaling, the 
segment with the lowest absorbance was presumed correct.  The lowest 
absorption that we can measure is the ``noise'' at the low point in $A$ 
near $\lambda$\,=\,2.5\,$\mu$m (4000\,cm$^{-1}$).  We therefore shifted the 
spectrum up to avoid negative values; the smallest $A$ is 0.001 or 0.0001, 
consistent with the signal-to-noise ratio of the spectral segment.  We 
scaled the thin film spectra to match the upturn at low frequency (high 
wavelength) for the slab sample.  We used the data over the Si-O stretching 
peak for the thin film, but otherwise used the thick film.  Spectra from 
the thick ($L \sim$0.47 mm) slab, rather than the ultrathin slab 
($L \sim$38\,$\mu$m), were used in the mid-IR due to noise considerations, 
although the data on the ultrathin slab were used to guide
merging.

For the reflectance spectra, scaling was not necessary; spectra were 
merged so that the noise was minimized.  For Fe, reflectivity is nearly 
100\% in the far-IR; attainment of this limiting value indicates that 
the absolute values were recorded.  For the ``cosmic silicate'' glass, 
the reflectivity in the visible is in accord with independent 
measurements of $n$.

\subsection{Optical function derivation}

We derived the optical functions for both ``cosmic silicate'' and metallic iron 
which are shown in Figures~\ref{optfunc1} and \ref{optfunc2}.

For the metallic iron spectra, we performed Kramers-Kronig analysis 
on the reflectivity data to obtain $n$ and $k$.  For details, see 
e.g., \citet{Spitzer62,Andermann65}
However, applying the standard methods for our
``cosmic silicate'' was problematic. Reflectivity $R$ for Mg-Ca-Si-O glasses is low
\citep[20\% at the peaks, compared to $\sim$80\% for 
crystalline silicates with similar composition; c.f.][]{Merzbacher}.
Consequently, the 
standard methods used to derive $n$ and $k$ for crystalline silicates 
sometimes fail to provide positive values for the dielectric function 
$\epsilon_{2}$ at all $\lambda$ for amorphous or glassy silicates.  
This problem can be understood in terms of glass structures.
Specifically, classical dispersion (damped harmonic oscillator model) 
assumes that the spectral peak breadths are due to damping 
(mode interactions), but in glasses, peak breadths also arise from a 
continuum of bond lengths in the glass and multiple overlapping peaks 
in the spectrum.  Likewise, Kramers-Kronig analysis fails to produce 
physically reasonable values when peaks are unresolved doublets 
\citep{giest02}.  

We use an alternative, more direct method to derive $n$ and $k$ values for our 
``cosmic silicate'' glass.  
Using absorption, reflectivity, and sample thickness, we determine the 
imaginary index of refraction $k(\lambda)$ via Eqs.~\ref{ascmeq}--\ref{klambda} 
and then use our reflectivity spectra $R$ to calculate $n(\lambda)$ using 
Eqs.~\ref{klambda}--\ref{Rfresnel}.

At all $\lambda$, $k$ was determined from absorption data where we 
corrected for reflections.  Let $\frac{I_{trans}}{I_{0}} 
\equiv$ transmittance, such that lab absorbance, in common logs is 
represented by Eq.~\ref{ascmeq},

\begin{equation}
\label{ascmeq}
a_{SCM} = - log_{10} ( I_{trans} / I_{0} ).
\end{equation}

For the polished slabs, we assumed that the measurement includes 
back reflections ($R$) from the slab's top and bottom faces and compute
the true or ideal absorption coefficient ($A$) after
\citet{hofm09} and \citet{Speck2014},

\begin{equation}
\label{Aeq}
A = 2.3026 [ a_{SCM} + 2 log_{10} ( 1 - R ) ] / L.
\end{equation}

For the ``cosmic silicate'' glass, Eq.~\ref{Aeq} overcorrects 
because the reflection from the back surface of the sample is less 
than $R$ due to light being attenuated slightly.  Thus, we adjusted 
the spectra manually for this effect and set $R = 0$ in Eq.~\ref{Aeq} 
for the ``cosmic silicate'' glass.
We then calculated the imaginary part of the complex refractive index $k$ using
Eq.~\ref{klambda}:

\begin{equation}
\label{klambda}
k_{\lambda} = \frac{A}{4\pi\nu}.
\end{equation}

The value of $k$ is fixed at 0 for $\nu =$0 cm$^{-1}$ and for very large $\nu$ above 
the X-ray region.

Laboratory reflectivity ($R$) equals $\frac{I_{meas}}{I_{0}}$.  Iron-free 
glasses are very transparent and $R$ is thus very low in the near-IR to 
visible; in this range, we computed $R$ from Fresnel's relationship, 
Eq.~\ref{Rfresnel},

\begin{equation}
R = [ ( n - 1 )^{2} + k^{2} ] [ ( n + 1 )^{2} + k^{2} ],
\label{Rfresnel}
\end{equation}

\noindent
using measured $n_{vis}$ and $k =$0.  At other wavelengths, we invert 
Eq.~\ref{Rfresnel} to provide $n(\lambda)$ from measured $R$ and $k$ 
(Eq.~\ref{nlambda}):

\begin{equation}
n(\lambda) = \frac{ 1 + R }{ 1 - R } + \sqrt{ -1 - k^{2} + ( \frac{ 1 + R}{ 1 - R } )^{2} }
\label{nlambda}
\end{equation}


For metallic iron, our $n$ and $k$ values extend to at $\lambda =$0.19\,$\mu$m; 
we did not extrapolate to shorter wavelengths.  
All metals have nearly 100\% reflectivity in the far-IR, which arises from the 
nearly free electron model of \citet{Drude} as discussed in \citet{Wooten}.

For the ``cosmic silicate'' glass, because the composition was nearly 
1\,MgO $+$ 1\,SiO$_2$, $n$ and $k$ were averaged from data on MgO and 
silica glass in \citet{Palik85}.
Also, $n$ is pinned at unity above the X-ray region 
and $k$ goes to 0 at very short wavelength.
%
At long wavelengths (far-IR), we assumed that $k$ decreases to 10$^{-7}$ at 0.001\,cm$^{-1}$ 
(10\,m} and that $n$ asymptotes to 3.16 at 0.001\,cm$^{-1}$
The value of $k$ is an estimate, based on the value of $k$ approaching 0 as the frequency goes 
to zero. 
The value of $n$ is from the low frequency dielectric function, calculated below.  

According to \citet{AppenBresker1952}, the dielectric constant is additive and can be 
estimated from the oxides using:

\[\epsilon = \Sigma \epsilon_i p_i\] 

\noindent
where $p_i$ is the fraction of the i$^{th}$ component
and $\epsilon_i$ was obtained at 0.45\,GHz (0.015\,cm$^{-1}$) 
from the table in \citet{AppenBresker1952}. 
See also the table on p. 318 of \citet{Scholze}\footnote{%
also available here: 
http://www.lehigh.edu/imi/docs$\_$GP/ Slides/GlassProp$\_$Lecture22$\_$Jain1.pdf page 24}.


\begin{figure}[!h]
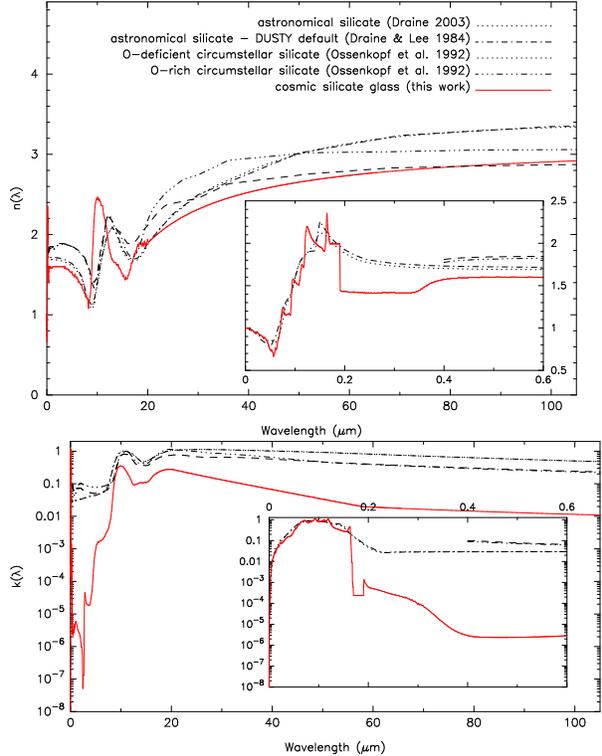

\includegraphics[angle=270, scale=.33]{f3a.eps}
\includegraphics[angle=270, scale=.33]{f3b.eps}
\caption{Comparison of complex refractive indices ($n$ \& $k$) for 
``Cosmic'' and ``Astronomical'' silicates. 
Solid line $\equiv$ this work; dotted and dashed lines $\equiv$ past works.
{\em Insets} are close-up views of the UV-vis region;
{\em Top}: real part of complex refractive index;
{\em Bottom}: imaginary part of complex refractive index;
$x$-axis is wavelength in $\mu$m. 
\label{optfunc0}}
\end{figure}

\begin{figure}[!h]
\includegraphics[angle=270, scale=.35]{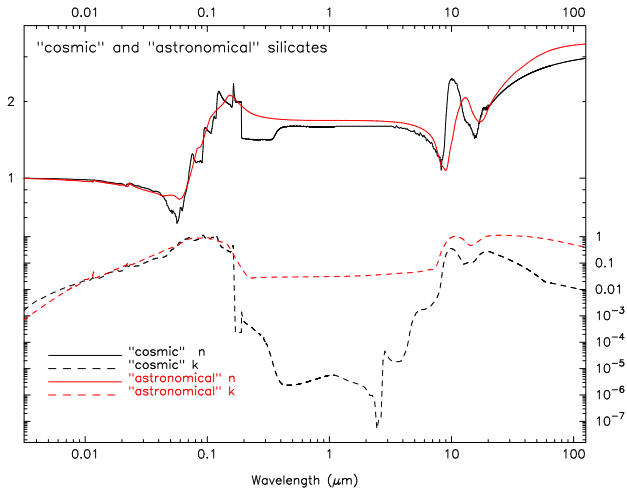}
\caption{New optical functions of ``cosmic silicate'' glass from this work 
(black lines), compared to ``astronomical silicate'' 
from \citet[red lines][]{draine03}.\label{optfunc1}}
\end{figure}

\begin{figure}[!h]
\includegraphics[angle=270, scale=.35]{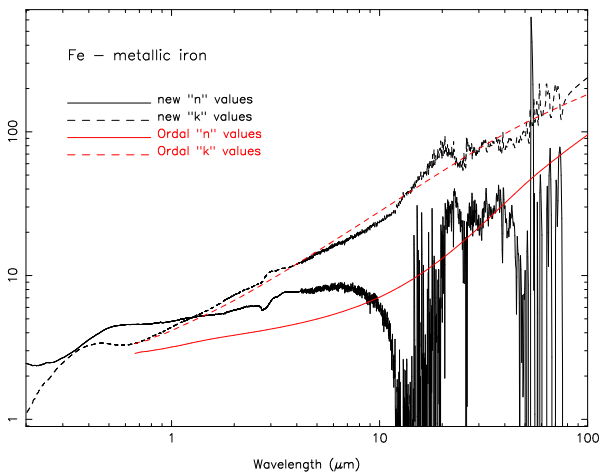}
\caption{New optical functions of metallic iron \label{optfunc2}
from this work (black lines), compared to values from
\citet{ordal} (red lines).}
\end{figure}

\section{Comparing optical functions}

Our new optical functions for ``cosmic silicate'' glass and metallic iron are compared to 
popular optical functions currently in use, i.e., \citet{draine03} for silicate and
\citet{ordal} for metallic iron in Figures~\ref{optfunc1} and \ref{optfunc2}.
Figure~\ref{optfunc1} clearly demonstrates that 
at $\lambda > 0.2$\,$\mu$m the differences 
are huge. In particular, the imaginary part of the complex refractive index ($k$) is several 
orders of magnitude lower in our new ``cosmic silicate'' glass measurements 
than in those derived by \citet{draine03}.
In part this difference comes from the difference in methods. The new data are 
determined from experimental laboratory data 
whereas the \citet{draine03} data are interpolated in this region and were specifically fixed 
to get high enough opacity to match observations. 

Figure~\ref{optfunc2} shows that our new optical functions for metallic iron match the 
overall strength and shape of the data from \citet{ordal}. However, our higher resolution
spectroscopy measurements reveal details with the optical properties of metallic iron not 
seen in the Ordal data. In particular the real part of the refractive index ($n$) drops to a 
very low value at $\lambda \sim$20\,$\mu$m and $\sim$\,50\,$\mu$m; meanwhile, 
the imaginary part, $k$, shows a bump around 30\,$\mu$m. 
In addition, the new optical functions extend down into the UV-Visible part of the spectrum, 
rather than ending at the NIR/visible boundary. 
This is very important for modeling in the interstellar medium or stars above about K type, 
because the majority of the flux is emitted in this higher energy region.

Our values differ from and better represent Fe than those of \citet{ordal} for the 
following reasons.  
The tables of $n$ and $k$ in \citet{ordal} are widely spaced (by 20\,cm$^{-1}$) 
in the far-IR; the spacing is even greater at shorter wavelengths.  
Also, the \citet{ordal} data were obtained from a Kramers-Kronig analysis that
incorporated a \citet{Drude} model at high and low frequencies (short and long wavelengths). 
We did not use a Drude model.  Kramers-Kronig analysis is effectively a smoothing function; 
this smoothing, combined with their low resolution, makes the \citet{ordal} $n$ and $k$ data 
very smooth and prohibits resolution of narrow features.  Our high-resolution iron data 
resolve the narrow features.  In the visible, we also see a feature at $\sim$0.4--0.5\,$\mu$m
in our $n$ and $k$ data that was observed in previous studies \citep[see e.g.][]{Palik91} 
with a discontinuity between their non-overlapping data sets.  
Our data are continuous across this feature and so improve upon these past works.  
In the near-IR, the small feature at 2--3\,$\mu$m in our iron data is not an artifact and 
is observed due to our high instrumental resolution. Therefore, we have provided improved 
$n$ and $k$ at higher resolution. The trade off is the noise in the far-IR longer than 
10\,$\mu$m; however, as will be seen in \S~\ref{sec:models} 
this noise does not affect our radiative transfer models of the observations.


\section{Radiative transfer modeling}
\label{sec:models}

\subsection{\it DUSTY}

\begin{figure*}[!htb]
\includegraphics[angle=270, scale=.65]{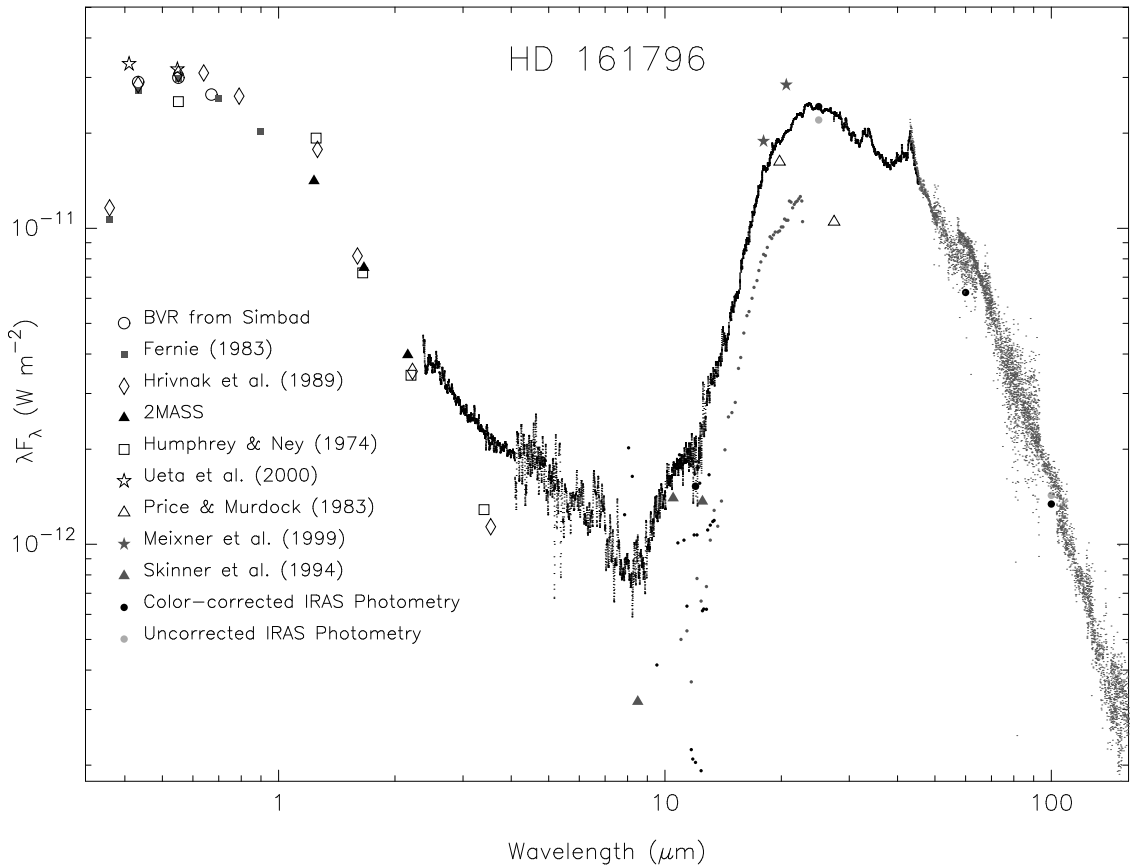}
\caption{Spectral Energy Distribution (SED) of HD\,161796.
Very small black dots are the {\it ISO} SWS data; 
very small grey dots are the {\it ISO} LWS data;
medium black filled circles are the short wavelength {\it IRAS} LRS data;
medium grey filled circles are the long wavelength {\it IRAS} LRS data;
large filled circles are the {\it IRAS} photometry data (grey are uncorrected; black are color corrected). Sources for other photometry points are given in the legend.
$x$-axis is wavelength in microns; 
$y$-axis is flux ($\lambda$\,$F_\lambda$) in W\,m$^{-2}$.
\label{Fig:SED}}
\end{figure*}

We used the 1-D radiative transfer program {\it DUSTY} 
\citep{Ivezic1995, Nenkova2000} to determine  how our new 
``cosmic silicate'' glass optical functions perform against previous 
``astronomical'' and ``circumstellar'' silicates $n$ and $k$. 
For our test case, we built simple models for HD\,161796, 
a dusty oxygen-rich pre-planetary nebula/post asymptotic giant branch star.
We chose HD\,161796 for several reasons: 
(1) it has been observed many times over several decades and over a 
large wavelength range, and thus has a well sampled spectral energy
distribution (SED), as shown in Figure~\ref{Fig:SED};
(2) its central star is hot enough to emit significant UV radiation, but not so hot that 
photo-dissociation and ionization play a major role in the nebula, 
simplifying the modeling; and 
(3) although not completely spherical, it is a 
remarkably round object \citep{ueta99}
and thus using a 1-d RT code is a reasonable approximation.

\subsection{Modeling HD\,161796 with synthetic optical functions}

\begin{figure}[!h]
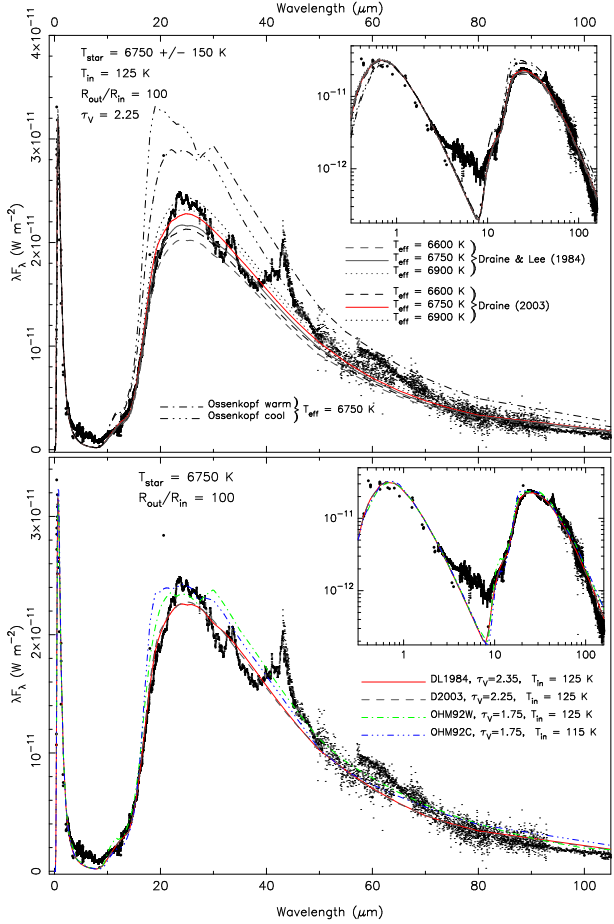

\includegraphics[angle=270, scale=.35]{f7a.eps}\\
\includegraphics[angle=270, scale=.35]{f7b.eps}
\caption{Best-fitting {\it DUSTY} models for ``Astronomical  Silicate''.
$x$-axis is wavelength in microns;
$y$-axis is flux ($\lambda$\,$F_\lambda$) in W\,m$^{-2}$.
Main panel is a linear-linear plot; inset is a log-log plot of the same data.
In all panels, large black filled circles = photometry points from Figure~1; 
very small dots = {\it ISO} spectrum.
{\em Top}: Best model fit is achieved using complex refractive index 
from 
\citet{draine03}.
The effect of varying $T_{\star}$ by $\pm$150K is negligible.
{\em Bottom}: Best-fitting {\it DUSTY} models using complex refractive index 
from \citeauthor{draine03,dl84,ohm92} for $T_{star} =$6750K and varying opacities.
\label{Fig:DUSTY1}}
\end{figure}

We have generated models using the new ``cosmic silicate'' glass $n,k$ data 
presented herein as well as using synthetic 
optical functions from \citet{draine03,dl84,ohm92}, which
are compared in Fig.~\ref{Fig:DUSTY1}.

For our models we adopt the central star temperature, $T_{\star}$ = 6750\,K, from 
\citet{hoozgaad}, 
which falls within the currently acceptable range for the object's central 
F3Ib class star \citep[6699\,K, 6600\,K, 6700\,K, 
6900--7350\,K;][respectively]{Kyotyukh,Leonid,Lang} and previous models 
\citep{skinner,meixner}. 
Fig.~\ref{Fig:DUSTY1} shows that the effect of varying this $T_\star$ 
value by $\pm$150\,K is negligible \citep[c.f.][]{DePew}.
DUSTY and many other RT models invoke Mie theory to calculate absorption and scattering 
cross-sections when a user supplies $n$ and $k$. 
The simple models shown here are made with the default DUSTY settings, 
which implicitly assume separate populations of spherical grains.
However, spherical grains are probably inappropriate in many astrophysical environments and 
are not the best representation of silicate features or of metals, 
whose FIR behavior is extremely shape and agglomeration dependent.  
For "real" models, we recommend supplying 
absorption and scattering coefficients for non-spherical grains calculated directly from  
$n$ and $k$. In that way, our new $n$ and $k$ 
values allow users to investigate the effect of grain size and shape
\citep[e.g.][]{pollack94, min}.

Figure~\ref{Fig:DUSTY1} ({\em top}) shows the effect of using the best-fitting model 
for the \citet{draine03} function model with the original ``astronomical silicate'' 
from \citet{dl84}, and with the warm and cold ``circumstellar silicate''
optical functions from \citet{ohm92}.
Figure~\ref{Fig:DUSTY1} ({\em bottom}) shows the best-fitting models using the 
original ``astronomical silicate'' optical functions from \citet{dl84} and those from
\citet{ohm92}, all three of which are embedded options in {\it DUSTY}. 
All model parameters are shown in Table~\ref{param1}.
The difference in performance between \citet{dl84} and \citet{draine03} is that the 
V-band optical depth is very slightly lower when using the newer \citet{draine03}
optical function.
Using the \citet{ohm92} synthetic functions requires even lower V-band optical depths.

\begin{table}
\caption{Best-fitting {\it DUSTY} models using synthetic optical functions from 
\citet{dl84,draine03,ohm92} \label{param1}}
%
\begin{tabular}{lcc}
\hline
source & $\tau_{V}$ & $T_{in}$ \\
\hline
D2003  & 2.25       & 125 \\
DL84   & 2.35       & 125\\
OHM92-C& 1.75       & 125\\
OHM92-W& 1.75       & 115\\
\hline
\end{tabular}
\end{table}

We assumed a radial dust density distribution of 
$\frac{1}{r^2}$ to reflect a constant mass-loss rate. 
We also initially assumed that the outer radius of the dust shell ($R_{\rm out}$)
is 100\,$\times$ the inner radius ($R_{\rm in}$), 
although the models are not very senesitive to this parameter.
\citet[][and references therein]{Speck09} demonstrated that the inherent 
degeneracy in RT modeling means that this number can be changed by compensating 
using the inner dust radius/temperature. Therefore we arbitrarily chose this 
value for comparison purposes.
We used an MRN grain-size distribution to be consistent with the assumptions of 
\citet{draine03,dl84,ohm92}. \citet{Speck09} showed that 
{\it DUSTY} models of spectral energy distributions are not very sensitive to the 
precise grains size  distribution as long as the distribution does not exclude many 
small grains or include very big grains ($>$1\,$\mu$m).

\subsection{Modeling HD\,161796 with our new ``cosmic silicate'' optical functions}

Our purpose in developing these simple models (an example of which is shown in Fig.~\ref{femodel})
is not to make an exhaustive study 
of the parameter space, or to optimize the fit to this particular object's SED, but instead to 
demonstrate the effect of using the new, real mineral 
data presented herein compared to more commonly use ``synthetic'' complex 
refractive indices. 
We have deliberately chosen the minimum number of dust components 
necessary for a fit and 
commonly used model parameters such as the 
$\frac{1}{r^2}$ radial density distribution and the MRN grain-size distribution with 
spherical grains.
The effect on radiative transfer modeling of choosing other parameters 
is discussed by \citet{Speck09}.

\subsubsection{``Cosmic silicate'' ONLY}

The initial models assumed a single dust component; later models included 
metallic iron and ``cosmic silicate''.
We also included external heating of the dust shell by 
the interstellar radiation field as well as a pile-up of material in the 
outer regions of the dust shell where the stellar outflow meets the 
interstellar medium in order to investigate whether composition or other 
factors 
improve the match to the shape of the SED \citep[c.f.][]{gil86,ypk93}.

Models using only our new ``cosmic silicate'' glass optical
functions do not match the data or the models using synthetic optical functions. 
There are two issues: 
(1) The shape of the SED produced using ``cosmic silicate'' is too symmetric and cannot match the observed emission at far-IR wavelengths; and
(2) the visual optical depth required is unfeasibly high ($100 < \tau_V < 150$).
We investigated the effect of varying several model parameters in an attempt to 
alleviate the 
far-IR problem including 
the radial density distribution, 
the geometrical size of the dust shell, 
inclusion of crystalline silicates,
inclusion of a step function in the density distribution to replicate the potential 
pile-up where the circumstellar materials meet the ISM,  and 
inclusion of 
external heating of the dust shell by the interstellar radiation field
\citep[see, e.g.,][]{gil86,ypk93}. None of these resolved either problem with the 
``cosmic silicate'' only models.

\subsubsection{``Cosmic silicate'' and metallic iron models}

As discussed above (\S~\ref{pastwork}), grains containing a mixture of metallic iron and 
iron-free silicates have long been invoked to account for the low opacity of silicates 
either explicitly 
\citep[e.g.][]{Merrill76,ohm92} or 
implicitly \citep[e.g.][and references therein]{draine03}.
Condensation models suggest that iron should form as metallic iron grains 
rather than as part of a solid solution of ferro-magnesian silicates 
\citep[e.g.][]{lf99}.
Therefore we also included metallic iron grains into our models, along with
``cosmic silicate'' glass, in varying proportions.
DUSTY assumes that the grains are co-spatial; i.e.\ the absorption and 
scattering cross-sections are calculated assuming the different grain compositions 
have the same size distribution and have the same temperature 
(essentially it a composite grain, comprising a silicate sphere and an iron 
sphere in thermal contact.)

\begin{figure}[!h]
\includegraphics[angle=270, scale=.35]{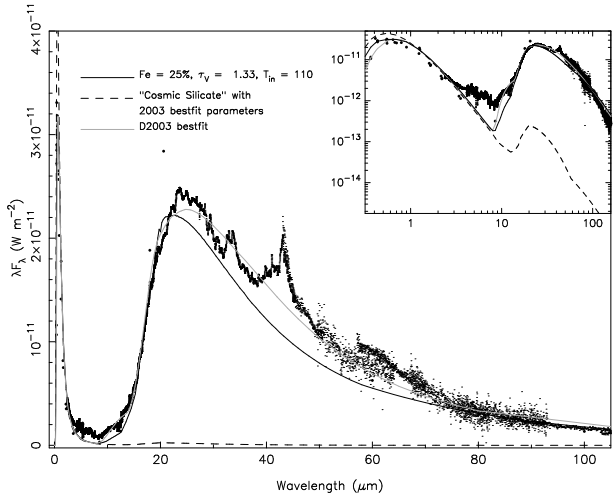}
\caption{Best-fitting {\it DUSTY} models for ``cosmic silicate'' 
with metallic iron included.
$x$-axis is wavelength in microns; 
$y$-axis is flux ($\lambda$\,$F_\lambda$) in W\,m$^{-2}$.
Main panel is a linear-linear plot; inset is a log-log plot of the same data.
In both panels, large black filled circles = photometry points from Figure 1; 
very small dots = {\it ISO} spectrum.
Thin black dashed line = model with ``cosmic silicate'' $n$ \& $k$ from this 
work, but \citet{draine03} best-fitting 
model parameters;
thin grey solid line = best-fitting model using \citet{draine03};
thick black solid line = best-fitting using ``cosmic silicate'' $n$ \& $k$ with 10\% metallic iron;
thick black dotted line =  best-fitting using ``cosmic silicate'' $n$ \& $k$ with 20\% metallic iron.
\label{femodel}}
\end{figure}

 Figure~\ref{femodel} shows that a modest addition of metallic iron dramatically 
improves both the shape and the optical depth requirements for the models. 
With only 25\% metallic iron grains, we can reduce the optical depth to 
$\sim$2, comparable to the models with synthetic optical functions.

Whereas including iron into the models improves the values for the visual optical depth, 
we still have a modeled SED that is narrower than either the observations or 
the models using synthetic optical functions. 
Inclusion of crystalline silicates 
\citep[which are obviously present based on the {\it ISO} spectrum 
in Fig.~\ref{Fig:SED} and have been included in previous 
models, e.g.][]{hoozgaad}
also does not alleviate the problem of SED shape and produces short wavelength 
crystalline features not seen in the observational data.

\subsubsection{Two-layer models}

\begin{figure}[!h]
\includegraphics[angle=0, scale=.166]{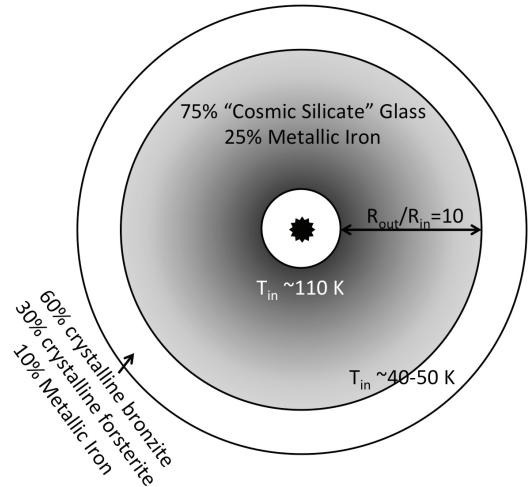}
\caption{Schematic cartoon of the 2-layer {\it DUSTY} models
\label{2layerschem}}
\end{figure}

\begin{figure}[!htb]
\includegraphics[angle=270, scale=.31]{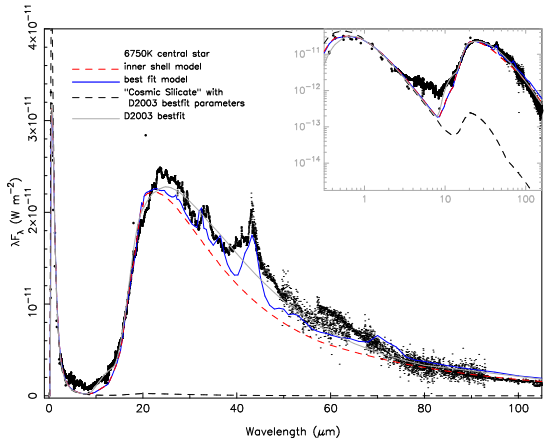}
\caption{Best-fitting {\it DUSTY} models using two-layer approach.
$x$-axis is wavelength in microns; 
$y$-axis is flux ($\lambda$\,$F_\lambda$) in W\,m$^{-2}$.
Main panel is a linear-linear plot; inset is a log-log plot of the same data.
In both panels, large black filled circles = photometry points from Figure~\ref{Fig:SED}; 
very small dots = {\it ISO} spectrum.
Thin black dashed line = model with cosmic silicate $n$ \& $k$, and
\citet{draine03} best-fitting 
model parameters;
thin grey solid line = best-fitting model using \citet{draine03};
thin red dotted line =  inner shell model
thick blue solid line = 2-layer model = inner model with thin shell of crystalline 
silicate and metallic iron on the outside. See schematic cartoon in Fig~\ref{2layerschem}. 
\label{fig:2layer}}
\end{figure}

Consequently, we used {\it DUSTY} to build a 2-layered model, which is shown schematically in 
Fig.~\ref{2layerschem}. The inner part of the shell is modeled to get the best fit shortward 
of 35\,$\mu$m. The output from this model is then the inner central source for the second layer 
model.
Because RT modeling is inherently degenerate, there are several models that 
produce good fits, summarized in Fig~\ref{2layerschem}. 
All good fits consist of an inner region dominated by ``cosmic silicate'' glass, 
with 25\% metallic iron, and a geometrically thin outer zone 
where the composition no longer contains any glassy silicate. Instead the thin outer layer 
contains 10\% metallic iron, 30\% Fe-rich pyroxene from \citet[bronzite;][]{henmutsch} 
and 60\% forsterite \citep[from][]{Pitman_etal2013}.
%
\citet{henmutsch} do not state whether the iron (FeO) 
content is given as mole \% or weight\%.  
However, in either case the sample should be referred to as Fe-rich enstatite 
rather than bronzite. We use this mineral as an example of a single crystalline 
silicate, rather than an attempt to match all the features.
The best 2 layer model is shown in Fig.~\ref{fig:2layer}.
 
This model may suggest that crystallization of silicates occurs either as a result of
shock heating of grains where the circumstellar medium ploughs into the surrounding ISM; 
or as a result of high energy (FUV) photons coming from the ISRF.
It is possible that this thin layer is also where many of the silicate grains are 
destroyed, which would account for the difference between circumstellar and 
interstellar silicate features. Indeed, it has been posited that most interstellar 
silicates form in cold, dark molecular clouds rather than being ejected from dying 
stars \citep[e.g.,][]{zhukovska,draine09}. However, this idea is controversial
\citep[see, e.g.][]{jonesnuth}.

\subsection{Iron optical functions}
There is a subtle effect as a result of replacing the widely used metallic 
iron optical functions from \citet{ordal} with our new metallic iron optical functions.
In order to match the model using the \citet{ordal} data we had to either increase the visual 
optical depth or increase the fractional abundance of metallic iron in the model.

\section{Discussion}
\label{sec:discuss}

It is clear from the way in which the optical functions for 
``astronomical silicates'' were derived that they have some limitations.
At the short wavelength end, the materials used do not actually represent the 
solids we expect to be dominating astrophysical environments. 
At the long-wavelength end, the data are derived from astronomical 
observations and assumptions and thus provide no flexibility to match 
environments with different dust components. 
Past attempts to derive the optical functions for astrophysically-relevant silicates
were hampered by the lack of availablity of consistent laboratory data at that time.
Decades later, we now have the means to overcome at least some of these 
issues.

The 10\,$\mu$m silicate feature occurs in many dusty astrophysical environments. 
It varies from object to object and even within a single object both temporally 
\citep[e.g.,][]{monnier98,GuhaNiyogi} and 
spatially \citep[e.g., in $\eta$ Car, N. Smith, Pers. Comm; and][]{GuhaNiyogi2}. 
Within a single type of astrophysical object, the feature shows 
huge variations in terms of its peak position, width and its ratio to the 
$\sim 18\mu$m feature \citep[e.g.,][]{speck98,ohm92}. 
Variations in feature or SED shape from object to object cannot be analyzed modeling with
synthetic optical functions.

Small circumstellar dust particles are likely much more compositionally and structurally 
inhomogeneous than the simple “glass + Fe” cases we have considered here.  Thus, we are 
providing data on well-characterized endmember glass to allow modelers to mix different 
compositions and impurities as they choose.  Our data also can aid in distinguishing between 
dust that is glassy, crystalline, or some combination of the two, which is important because the 
degree of structural homogeneity has implications for its formation, and subsequent processing, 
evolution and destruction \citep{SWH11}.


\subsection{Comparison to previous models}

HD\,161796 is a well-studied object and many previous models exist, 
which makes it an excellent test case.
\citet{hrivnak} produced models of several candidate pre-planetary nebulae, 
including HD\,161796, using dust opacities derived from \citet{vk88}. 
Their output parameters depend on distance, but are consistent with the 
later models in terms of inner dust radius and stellar luminosity.

\citet{justtanont} made a very simple model of HD\,161796's SED using a 
6000\,$L_\odot$ F-type star surround by dust with a single blackbody 
at 90\,K modified by the emissivity from disordered olivine from \citet{kh79} 
and assuming an MRN size distribution.
Their best-fitting model did not match the SED at all wavelengths, and although it was somewhat 
improved by the addition of magnetite, there remained mismatches between the model and 
the observed SED.

\citet{skinner} used multiwavelength imaging to constrain their models of HD\,161796.
Their model assumed an effective star temperature of 6300\,K and used astronomical 
silicate from \citet{dl84} in an MRN distribution. Their best-fitting model gave a 
stellar luminosity of 3600$\L_\odot$ and an inner dust radius of 8.9$\times$10$^{15}$\,cm.

Up to this point, the past models assumed spherical symmetry (as do we). 
Furthermore, there was not yet observational evidence for crystalline silicates.
With the advent of {\it ISO}, more detail in the far-IR spectrum was revealed and led 
to modeling including crystalline silicates.
\citet{hoozgaad} modelled the full 
2--200\,$\mu$m {\it ISO} SWS $+$ LWS spectrum together with additional data extending the 
SED from the optical to the millimeter wavelength range.
Their model assumed a star of 3000$\L_\odot$ at 6750\,K 
with an inner dust radius of 21$\times$10$^{15}$\,cm 
(and outer radius of 63$\times$10$^{15}$\,cm). 
\citet{hoozgaad} developed a two layer model in which the 
inner dust is dominated by amorphous silicate and the outer dust includes a mantle of 
water ice. They 
used a number of different optical functions in their model in order to cover the full 
wavelength range, but the amorphous silicate was solely from \citet{ohm92} and was 
extrapolated out to the extremes of the wavelength range. In addition, they 
included crystalline water ice, as well as crystalline silicates in the form of 
enstatite and forsterite. They focussed on developing two parts of their model - 
the core-mantle grains and the crystalline silicate fraction, although they found that
the lack of sensitivity of the 
models to the precise abundances of crystalline silicates made 
abundance constraints weak, but found that $\sim$10\% by mass was reasonable. The average 
temperature for the amorphous silicate grains was found to be 110\,K 
(and ranged from 47--137\,K) while the crystalline grains were cooler 
(at $\sim$50-90\,K), consistent with our results.

\citet{meixner} produced a 2-d model of HD\,161796 for which they combined both 
multiwavelength imaging and the SED to constrain their models.
Their best-fitting model had a 2800$\L_\odot$ 
star with an effective temperature of 7000\,K. 
They found an inner dust temperature of 110\,K and a corresponding inner dust radius of
11$\times$10$^{15}$\,cm. Whereas the model is axisymmetric, the change in dust density 
from equator to pole is small.
Their model included both forsterite and enstatite but the coarse 
spectral resolution precluded fitting the narrow crystalline silicate features in detail.
The grain size distribution is KMH-like with a higher upper limit on grain size, 
and they assumed the dust composition from \citet{hoozgaad}.

Given the range of optical functions and model types used, it is interesting 
(and reassuring) that all models including ours end up being dominated by amorphous silicate 
dust at $\sim$100\,K. However, previous models used synthetic optical 
functions for amorphous silicates and thus did not 
include the correct Fe/Mg fraction and cannot reveal the shell structure seen in our 
modeling results.

\section{Conclusions}
\label{sec:conclu}

We have presented new optical functions
for iron-free glassy silicate with chondritic composition and metallic iron,
extending from UV to far-IR wavelengths (0.19-250\,$\mu$m), which 
are measured directly in the laboratory 
using the same sample throughout. 
The ``cosmic silicate'' data extend to the X-ray region.
These new optical functions improve upon prior synthetic optical functions 
for ``astronomical'' and ``circumstellar'' silicates by
preserving the chemical composition, sample properties and the method of
optical function derivation (all lab data).
A simple model excursion using HD\,161796 as an example shows that:

\begin{enumerate}

\item{using synthetic optical functions automatically fits most SED slopes and thus 
provides false parameters for models}.

\item{inserting back-calculated $n$ and $k$ from observed data is not necessary to 
achieve the desired SED slope; only a small amount of metallic iron is needed to 
make the models sufficiently opaque};

\item{the precise ratio of iron/cosmic silicate  strongly affects the far-IR emissivity 
law. Thus, our new laboratory-derived optical functions can be used to determine 
how composition affects observed long-wavelength 
emissivities and emissivity indices ($\beta$ value), 
which is relevant to understanding many long wavelength 
observations including those from
{\it ALMA} and {\it Herschel}, {\it Spitzer}, {\it IRAS} and {\it ISO}};

\item{our models of HD\,161796 hints at crystallization of silicates at the outer edge 
of the circumstellar shells. While this is a preliminary result that requires further study,
this suggestion is not apparent when using synthetic optical functions};


\end{enumerate}

Because our cosmic silicate has chondritic abundance, it may not be precisely 
the composition of glassy silicates formed away from the solar neighborhood. 
Future work will include other silicate glass compositions including both iron-free and iron-bearing compositions.

\acknowledgments

This material is based upon work supported by the National Science Foundation under
Grant No. AST\,1009544.
The authors thank A. Buffard for preliminary {\it DUSTY} models,
A. Whittington for producing the synthetic glass samples 
and A. Corman  \& J. Goldsand for laboratory assistance.
Electronic tables of the spectra shown in this work are available 
as an electronic supplement to this article and at
\url{http://galena.wustl.edu/$\sim$dustspec/idals.html}.

\end{document}